\documentclass[conference]{IEEEtran}
\IEEEoverridecommandlockouts
\usepackage{cite}
\usepackage{amsmath,amssymb,amsfonts}
\usepackage{algorithmic}
\usepackage{graphicx}
\usepackage{textcomp}
\usepackage{xcolor}
\usepackage{booktabs}
\usepackage{multirow}
\def\BibTeX{{\rm B\kern-.05em{\sc i\kern-.025em b}\kern-.08em
    T\kern-.1667em\lower.7ex\hbox{E}\kern-.125emX}}
\begin{document}

\title{Edge-based Parametric Digital Twins for Intelligent Building Indoor Climate Modeling
\thanks{This study is partially funded by the Swedish Energy Agency (Energimyndigheten) and partially funded by the Swedish Innovation Agency (Vinnova).}
}


\author{\IEEEauthorblockN{Zhongjun Ni}
\IEEEauthorblockA{\textit{Department of Science and Technology} \\
\textit{Link\"oping University, Campus Norrk\"oping}\\
Norrk\"oping, Sweden \\
zhongjun.ni@liu.se} \\
\IEEEauthorblockN{Magnus Karlsson}
\IEEEauthorblockA{\textit{Department of Science and Technology} \\
\textit{Link\"oping University, Campus Norrk\"oping}\\
Norrk\"oping, Sweden \\
magnus.b.karlsson@liu.se}
\and
\IEEEauthorblockN{Chi Zhang}
\IEEEauthorblockA{\textit{Department of Computer Science and Engineering} \\
\textit{University of Gothenburg}\\
Gothenburg, Sweden \\
chi.zhang@gu.se} \\
\IEEEauthorblockN{Shaofang Gong}
\IEEEauthorblockA{\textit{Department of Science and Technology} \\
\textit{Link\"oping University, Campus Norrk\"oping}\\
Norrk\"oping, Sweden \\
shaofang.gong@liu.se}
}

\maketitle

\begin{abstract}

Digital transformation in the built environment generates vast data for developing data-driven models to optimize building operations. This study presents an integrated solution utilizing edge computing, digital twins, and deep learning to enhance the understanding of climate in buildings. Parametric digital twins, created using an ontology, ensure consistent data representation across diverse service systems equipped by different buildings. Based on created digital twins and collected data, deep learning methods are employed to develop predictive models for identifying patterns in indoor climate and providing insights. Both the parametric digital twin and deep learning models are deployed on edge for low latency and privacy compliance. As a demonstration, a case study was conducted in a historic building in Östergötland, Sweden, to compare the performance of five deep learning architectures. The results indicate that the time-series dense encoder model exhibited strong competitiveness in performing multi-horizon forecasts of indoor temperature and relative humidity with low computational costs.  

\end{abstract}


\section{Introduction}
Buildings are becoming smarter by integrating advanced information and communication technologies. Internet of Things (IoT) devices continuously collect data from buildings. As estimated by Statista~\cite{smart_buildings_data_online_2024}, 37 zettabytes of data were collected globally in 2020. These data provide opportunities for developing data-driven models to optimize building operations and equipment control~\cite{zhang_review_2021}. Cloud computing, with its enhanced computing and storage capabilities~\cite{liu_methodology_2021}, facilitates the handling of enormous IoT data. Furthermore, continuously collected data enables the creation of digital twins. A digital twin of a building models essential properties and relationships of interested physical entities~\cite{ni_enabling_2022}, ranging from the entire building to specific service systems like heating, ventilation, and air-conditioning (HVAC) systems~\cite{hosamo_digital_2023}. On the one hand, a digital twin reflects the latest status of its physical counterpart with ingested real-time data, which enables monitoring~\cite{ni_sensing_2021} and defect detection~\cite{lu_digital_2020}. On the other hand, integrated with advanced data analytics represented by deep learning-based predictive models, a digital twin can predict future conditions, which allows predictive maintenance~\cite{zhang_deephealth_2021} and energy optimization~\cite{ni_leveraging_2023}.

A suitable indoor climate is crucial for human comfort. Even for buildings equipped with modern HVAC systems, inappropriate operation of these systems might lead to thermal comfort issues~\cite{bakhtiari_evaluation_2020}. Moreover, this concern is relevant for historic buildings, where indoor climate directly affects heritage conservation~\cite{leijonhufvud_standardizing_2018}. While various studies~\cite{ni_enabling_2022, ni_improving_2021, khajavi_digital_2019, zhang_automatic_2022} have explored the creation of digital twins to enhance building operation and maintenance, many of them are cloud-centric. Relying on cloud computing could bring problems like bandwidth bottlenecks and data privacy concerns~\cite{hua_edge_2023}. The emergence of edge computing is intended to address these issues by moving computational data, applications, and services closer to the network edge, potentially enabling low-latency IoT solutions~\cite{wang_convergence_2020}. Although edge-based digital twins have been applied for job shop scheduling~\cite{wang_edge_2023}, computation state estimation~\cite{zhang_digital_2023}, and mobility services~\cite{wang_mobility_2022}, their application in the built environment is limited.

This study aims to integrate edge computing, digital twins, and deep learning for intelligent building indoor climate modeling. The solution is edge-centric, where both digital twin and deep learning-based predictive models are deployed on edge for low latency and privacy compliance. In contrast to 3D geometric models, parametric digital twins are created to capture evolving essential properties of buildings, e.g., various indoor environmental parameters. Instead of using computational fluid dynamics-based methods, deep learning methods are employed to develop predictive models for indoor climate. This solution can further incorporate advanced control strategies, such as model predictive control~\cite{drgoňa_all_2020}, to determine optimal control inputs for HVAC systems. As main contributions, this work:
\begin{itemize}
  \item Proposed an integrated edge-centric solution for intelligent building indoor climate modeling, serving as a valuable alternative to cloud-centric approaches.
  \item Conducted a thorough case study to assess the performance of five deep learning architectures, i.e., long short-term memory (LSTM), temporal convolutional network (TCN), temporal fusion transformers (TFT), neural hierarchical interpolation for time series (N-HiTS), and time-series dense encoder (TiDE), in predicting indoor climate. 
\end{itemize}

Besides the built environment, the proposed solution can also be adapted to manufacturing processes. For example, creating digital twins of manufacturing facilities enables proactive maintenance by predicting potential equipment failures. Additionally, equipment operations could be scheduled based on demand forecasts, and energy-saving measures might be adopted during periods of low production activity.

\section{Related Work}
\label{sec:related_work}

Modeling building indoor climate presents unique challenges, such as the complexity of capturing various influencing factors. As indoor climate measurements are typically represented as time series, recent advances in deep learning-based time series forecasting are first presented. Afterward, applications of intelligent edge computing and digital twins in the built environment are reviewed.



\subsection{Deep Learning for Time Series Forecasting}

Several deep learning architectures have been designed to tackle diverse time series forecasting problems across multiple domains~\cite{zhang_pedestrian_2023, zhang_spatial_2023, ni_study_2024}. These architectures include various types of neural networks, such as recurrent neural networks (RNNs) represented by LSTM~\cite{hochreiter_long_1997} and GRU~\cite{cho_properties_2014}, convolution-based known as TCN~\cite{bai_empirical_2018}, and attention mechanism-based like TFT~\cite{lim_temporal_2021}. In addition, some architectures, e.g., N-HiTS~\cite{challu_nhits_2023}, are based on a deep stack of fully connected layers.

Some studies challenged the validity of Transformer-based solutions for long-term time series forecasting tasks since time series data typically lack semantics~\cite{zeng_are_2023}. For example, Das et al.~\cite{das_long_2023} proposed TiDE based on dense multi-layer perceptrons. Although without recurrent, convolutional, or self-attention mechanisms, TiDE outperforms existing neural network-based models on time series forecasting benchmarks.

Much research has reported using some of these architectures to develop models for performing forecasting tasks in buildings, such as energy consumption prediction~\cite{ni_leveraging_2023, ni_study_2024}. However, a comprehensive comparison of which architecture is more effective in building indoor climate forecasting is lacking. Furthermore, the computational cost of deploying predictive models developed based on these architectures on edge devices has not been well studied.

\subsection{Intelligent Edge Computing and Digital Twins in the Built Environment}

Many studies have reported applying intelligent edge computing in the built environment, such as indoor localization~\cite{liu_edge_2017}, home intrusion monitoring~\cite{dhakal_machine_2017}, and resource allocation~\cite{singh_gru_2024}. The utilized inference methods include Bayesian theory~\cite{liu_edge_2017}, k-nearest neighbors algorithm~\cite{dhakal_machine_2017}, and RNNs like GRU~\cite{singh_gru_2024}. The aforementioned deep learning architectures, such as TFT and TiDE, have not been fully exploited.

Some work attempted to improve the operation and maintenance of buildings through digital twins. Typical application scenarios include anomaly detection for building assets~\cite{lu_digital_2020}, energy optimization~\cite{ni_improving_2021, hosamo_digital_2023, ni_leveraging_2023}, and building conservation~\cite{zhang_automatic_2022, ni_enabling_2022}. Although some of these studies claimed to perform preventive maintenance~\cite{khajavi_digital_2019, ni_enabling_2022}, predictive models were not integrated in digital twins. There is also a lack of integration between edge computing and digital twins in the built environment. Most of the applications in these studies are cloud-centric, i.e., digital twin models are deployed in the cloud, and related data are collected and then entirely uploaded to the cloud to update model status and for further analytics.

\section{Methodology}
\label{sec:methodology}

This section first describes the overall architecture of proposed solution. Then, methods for creating parametric digital twins and developing deep learning-based indoor climate predictive models are further explained separately.

\subsection{System Architecture}

\begin{figure}[!b]
\includegraphics{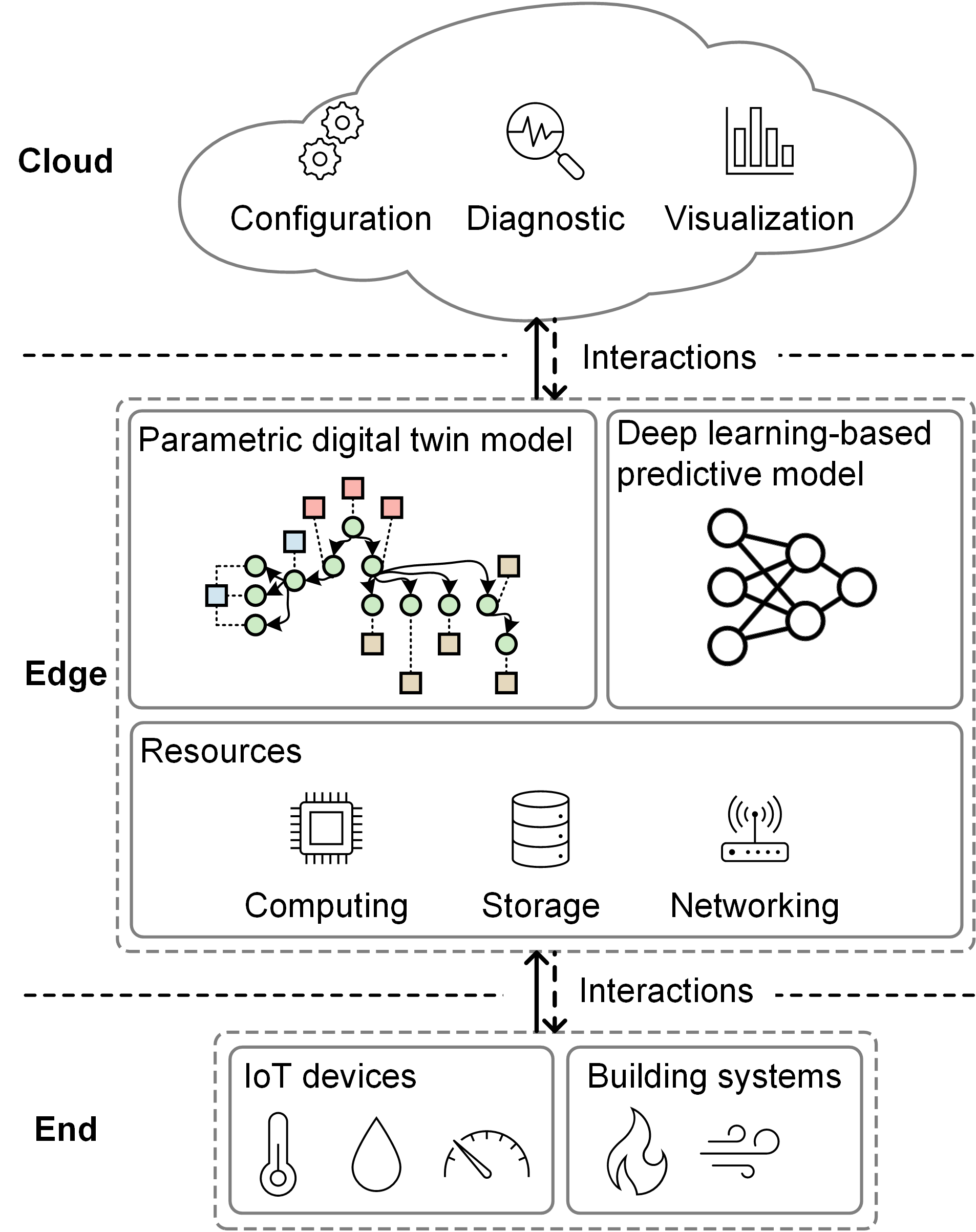}
\centering
\caption{An illustration of the architecture of proposed solution.}\label{fig:fig1}
\end{figure}

As depicted in Fig.~\ref{fig:fig1}, the architecture of the solution consists of three layers. The hierarchy from bottom to top is End, Edge, and Cloud. Each layer has its own responsibilities.

The end layer serves two main purposes. The first is consistently collecting data from a building through various IoT devices, such as sensors, actuators, and smart meters. The second is receiving commands from the edge or cloud and taking corresponding actions, such as adjusting set points of building systems like heating and ventilation.

The edge layer plays a centric role in the solution. It has computing, storage, and networking resources for performing tasks ofﬂoaded by the cloud. Collected data are analyzed directly on the edge instead of being entirely uploaded to the cloud. Both parametric digital twin and predictive models are deployed on the edge layer. The status of the digital twin model is consistently updated according to collected data. The predictive model queries data from the digital twin model as input and generates forecasts for guiding indoor climate control, such as heating, to keep indoor climate within the required bounds.

The cloud layer mainly provides access for configuring and managing the edge layer as well as user interfaces such as data visualization. For example, developed predictive models can be deployed to the edge through the cloud. As data analytics are offloaded to the edge, only a few data, e.g., insights generated from the edge, need to be uploaded to the cloud for further diagnostic. 

\subsection{Creation of Parametric Digital Twins}

A parametric digital twin model encapsulates essential information about physical entities. It also provides necessary reading or writing application programming interfaces (APIs) for updating model status from data sources and offering data access for analytics. As in our early studies~\cite{ni_enabling_2022, ni_leveraging_2023}, ontology was employed to ensure consistent data representation across heterogeneous building systems. The Brick ontology~\cite{balaji_brick_2018} was extended to support more types of sensors and outdoor weather conditions. Ontological modeling enhances the adaptability of the solution to other buildings.

\subsection{Development of Deep Learning-based Indoor Climate Predictive Models}

\begin{figure}[!tb]
\includegraphics[width=\columnwidth]{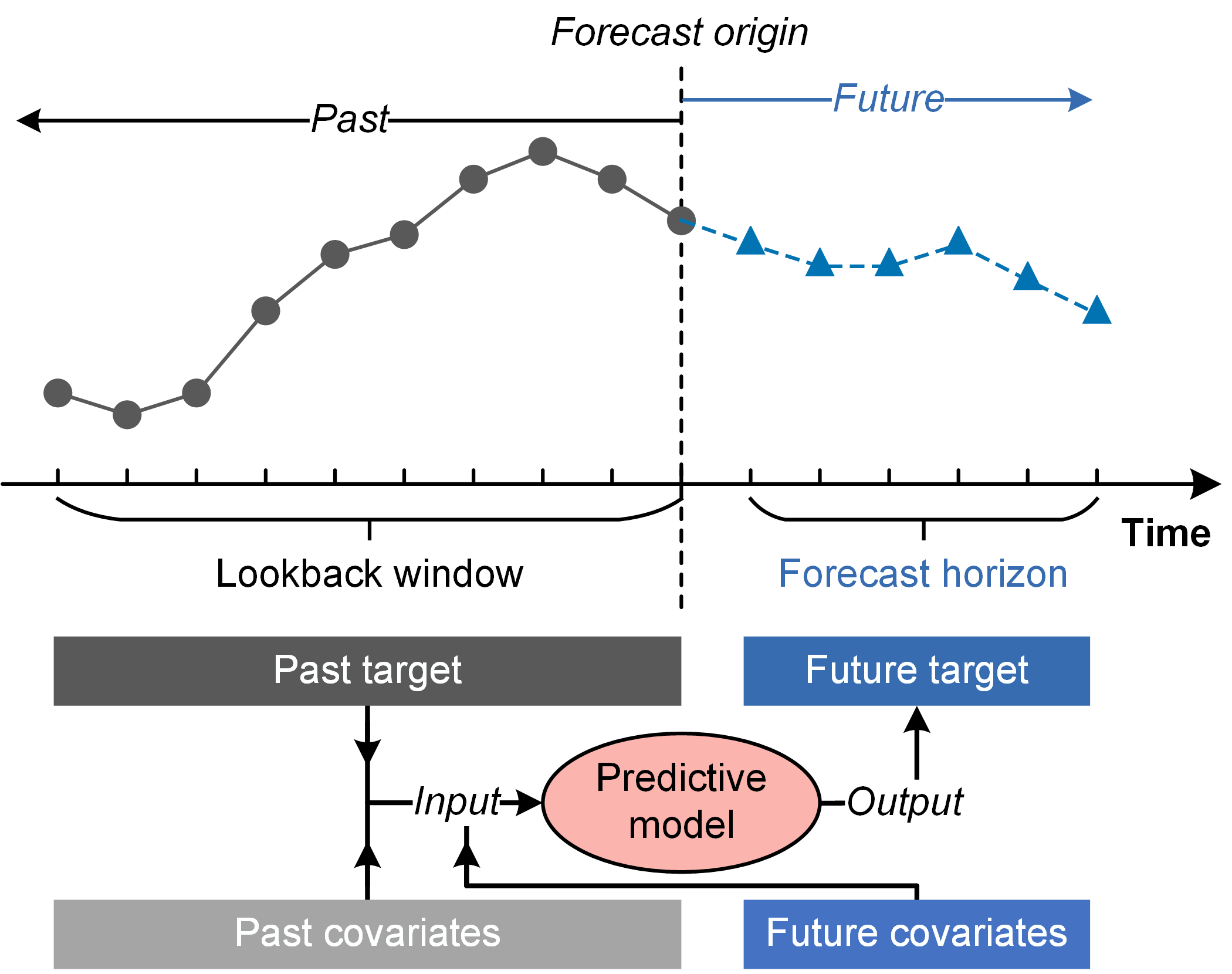}
\centering
\caption{An illustration of multi-horizon forecasting of building indoor climate. In this instance, black dots represent observed values of a particular environmental parameter over a lookback window in the past. Blue triangles are forecasts of the environmental parameter over a forecast horizon in the future. A predictive model uses past observations of a target variable and covariates over the loopback window, along with observations of covariates over the forecast horizon, to make predictions for the target variable.}\label{fig:fig2}
\end{figure}

This study aims to develop multi-horizon predictive models for building indoor climate. As shown in Fig.~\ref{fig:fig2}, given a forecast origin, a predictive model utilizes past observations of a target variable and covariates affecting it over the lookback window, along with observations of covariates over the forecast horizon, to generate forecasts for the target variable. Typically, the forecast horizon is set to 5--48 hours~\cite{afram_theory_2014}.

Predictive models were trained on a training set to minimize the total squared error. The performance of developed models was compared through two aspects: computational cost and prediction accuracy. The computational cost was expressed as inference time of a model on the edge in milliseconds. The inference time should generally be no longer than a few seconds for a feeling of responsiveness. The prediction accuracy was evaluated by two scale-independent metrics, namely coefficient of variation of the root mean square error (CV-RMSE) and normalized mean bias error (NMBE), over the entire test set. As in study~\cite{o_predicting_2019}, NMBE $\pm$5\% and CV-RMSE $\leq$~20\% are adopted as criteria. These two metrics are calculated with (\ref{eqn:cv-rmse}) and (\ref{eqn:nmbe}).
\begin{equation}
CV\textrm{-}RMSE=\frac{\sqrt{\frac{1}{n}\sum\limits _{t=1}^{n}(\hat{y}_{t} -y_{t})^{2}}}{\overline{y}} \times 100, \label{eqn:cv-rmse}
\end{equation}
\begin{equation}
NMBE=\frac{\frac{1}{n}\sum\limits _{t=1}^{n}(\hat{y}_{t} -y_{t})}{\overline{y}} \times 100,  \label{eqn:nmbe}
\end{equation}
where $n$ denotes the size of forecast horizon, $y_{t}$ is the actual value of a target variable at time $t$, $\hat{y}_{t}$ is the predicted value of the target variable at time $t$, and $\overline{y}$ is the mean actual value of the target variable over the forecast horizon.

\section{Case Study}
\label{sec:case_study}

To evaluate the performance of different deep learning architectures, a case study was conducted to develop predictive models for indoor temperature and relative humidity of a historic building.

\subsection{Description of Löfstad Castle}

\begin{figure}[!tb]
\includegraphics{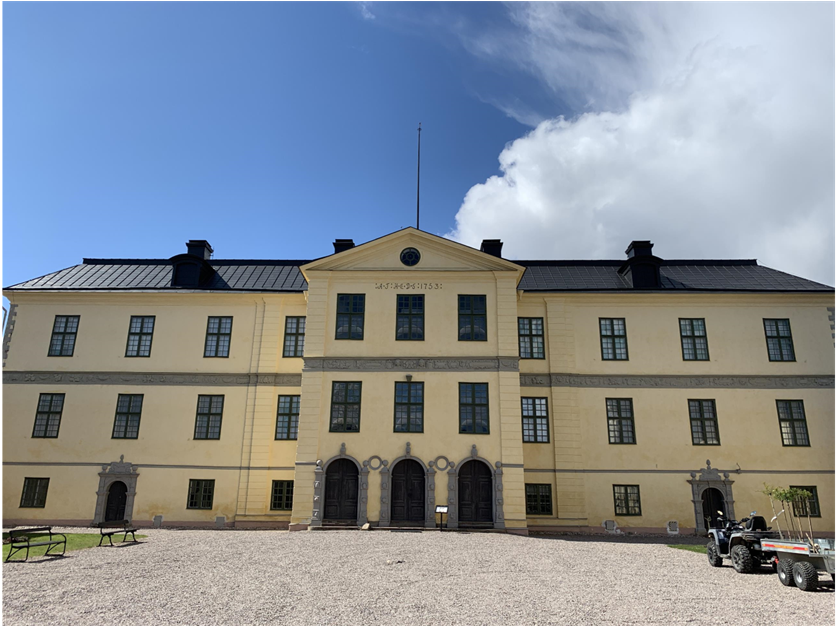}
\centering
\caption{Main building of the Löfstad Castle, Östergötland, Sweden.}\label{fig:fig3}
\end{figure}

Löfstad Castle is a listed historic building in Östergötland, Sweden, which has high cultural and historical values~\cite{lofstad_castle_online}. Today, the castle is open as a museum to visitors. The main building of the castle (see Fig.~\ref{fig:fig3}) has several floors, including a basement, three floors above the basement (ground, first, and second floors), and an attic. Currently, the main building is naturally ventilated, and only the first floor has employed electric radiant heating during the season when heating is needed. The building and collections are suffering from an inappropriate indoor climate. For instance, some wooden artifacts housed in rooms on the ground floor have been damaged by insect infestation. Parametric digital twins and indoor climate prediction models can provide insights into improving indoor climate to mitigate these deteriorations.

\subsection{Data Collection}

The dataset used for developing predictive models includes two parts: indoor environmental data and outdoor meteorological data. Indoor environmental data are collected through prototype sensor boxes made by ourselves, each containing multiple sensing devices and an edge platform. Among them, six different types of sensing devices are used for measuring seven environmental parameters.
\begin{itemize}
  \item AHT20 (Adafruit, New York, USA) measures temperature and relative humidity.
  \item MH-Z16 (Winsen Electronics Technology, Zhengzhou, CHN) measures CO\textsubscript{2} concentration.
  \item PPD42NS (Shinyei Corporation, New York, USA) detects particulate matter with a diameter greater than 1~$\mu$m.
  \item LM358 (Seeed Technology, Shenzhen, CHN) detects the relative intensity of ambient noise.
  \item TSL2591 (Adafruit, New York, USA) measures light intensity.
  \item UR18.DA0.2-UAMJ.9SF (Baumer, Frauenfeld, Switzerland) measures groundwater level.
\end{itemize}

\begin{figure}[!tb]
\includegraphics{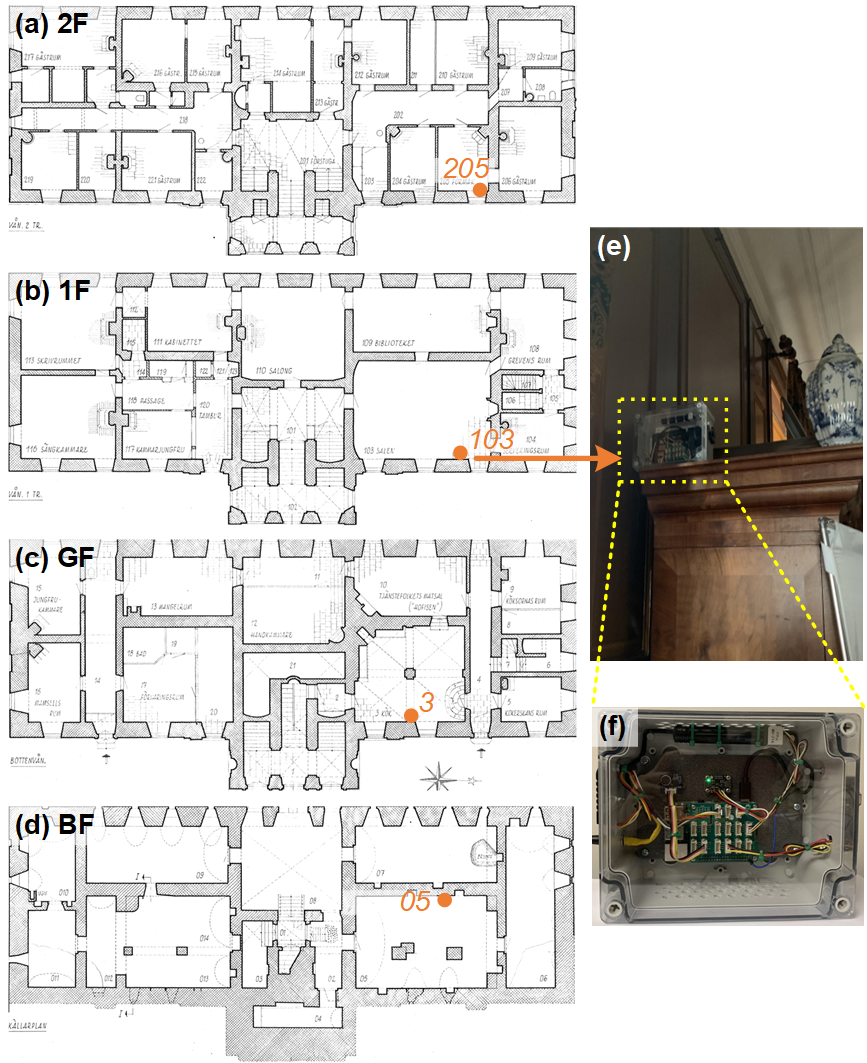}
\centering
\caption{Four rooms are selected for comparison. (\textbf{a}) Room 205 on the second floor (2F), (\textbf{b}) Room 103 on the first floor (1F), (\textbf{c}) Room 3 on the ground floor (GF), and (\textbf{d}) Room 05 on the basement floor (BF). The prototype sensor box deployed in Room 103 is depicted in (\textbf{e}). Each of the other three rooms also has been deployed a sensor box that looks like in subfigure (\textbf{f}). Orange dots mark deployment positions. Among the four rooms, only Room 103 has employed electric radiant heating.}\label{fig:fig4}
\end{figure}

The prototype edge platform includes a single-board computer, i.e., Raspberry Pi 4 Model B (Raspberry Pi Foundation, Cambridge, GBR), and a Grove Base Hat (Seeed Technology, Shenzhen, CHN). A 4G USB modem E3372-325 (Huawei, Shenzhen, CHN) is added for deployment locations lacking Wi-Fi connection to allow Internet access for the computing module. Except for the groundwater level sensor, the other five sensing devices and the edge platform are packaged in a plastic box (see Fig.~\ref{fig:fig4}{f}). A total of 13 sensor boxes are distributed throughout the main building from the basement to the attic, continuously monitoring the indoor environment on all floors. In this study, four rooms on different floors are selected for comparison (see Fig.~\ref{fig:fig4}{a} to~\ref{fig:fig4}{d}). The start date of data collection is February 13, 2023 for Room 05 and January 12, 2023 for the other three rooms.

Each environmental parameter is collected every 30 seconds, allowing for capturing rapid fluctuations. Downsampling is available for subsequent data analytics and visualization. The acquired large amount of multi-parametric data enables the creation of a parametric digital twin of the castle, providing a deeper understanding of the indoor climate and improving heating and ventilation strategies.

The meteorological data include hourly dry-bulb temperature, relative humidity, dew point temperature, precipitation, wind speed, wind direction, and global irradiance. They are obtained through open APIs provided by the Swedish Meteorological and Hydrological Institute. The global irradiance is collected according to the latitude and longitude of Löfstad Castle, while other meteorological data are from a weather station located around 7~km away from the castle.

\begin{figure}[!b]
\includegraphics[width=\columnwidth]{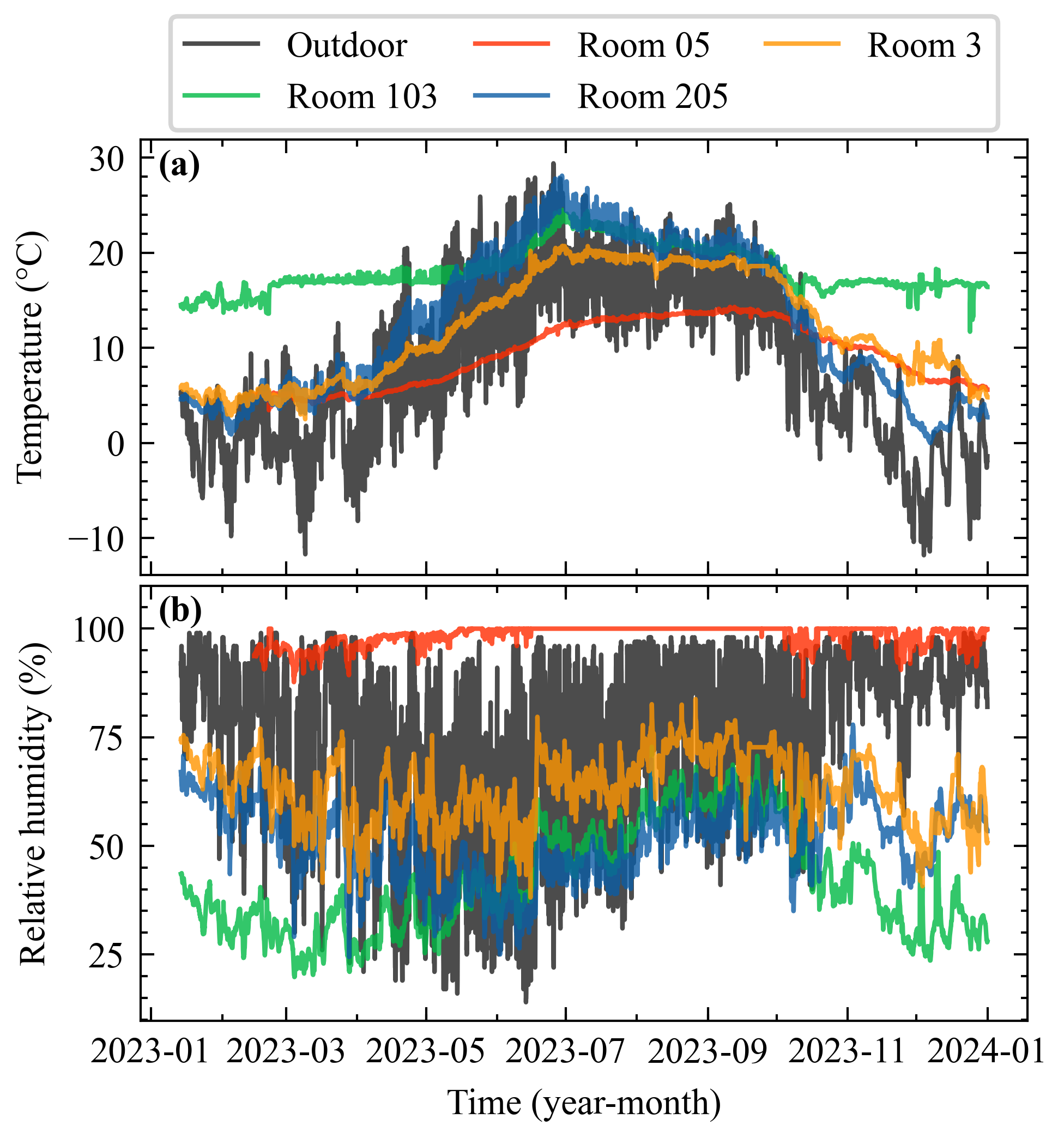}
\centering
\caption{Historical hourly (\textbf{a}) temperature and (\textbf{b}) relative humidity of outdoor and four selected rooms from a beginning time to 23:00 on December 31, 2023. The beginning time is 00:00 on February 14, 2023, for Room 05 and 00:00 on January 13, 2023, for the other three rooms and outdoor. Hours appearing in this paper are expressed in 24-hour format and are all in the timezone of Central European Time.}\label{fig:fig5}
\end{figure}

Historical hourly outdoor dry-bulb temperature and relative humidity as well as indoor temperature and relative humidity are shown in Fig.~\ref{fig:fig5}. There are some correlations between outdoor and indoor temperatures and relative humidity. As depicted in Fig.~\ref{fig:fig5}{a}, temperatures in Rooms 05, 3, and 205 were mainly determined by outdoor temperature throughout the year but had fewer hourly fluctuations over days. This phenomenon is due to the fact that the massive walls of Löfstad Castle acted as a thermal buffer to stabilize indoor temperature, especially for Room 05 on the basement floor. In the non-heating season (e.g., from June to September), temperature in Room 103 was also affected by outdoor temperature. However, in the heating season, temperature in Room 103 remained stable around 17~$^{\circ}$C due to the work of the electric radiant heating system. Similar correlations can be found between outdoor and indoor relative humidity. Room 05, situated in the basement with tiled floors directly laid on the soil, experienced constant moisture evaporation into the basement, maintaining the relative humidity above 90\% for most of the time.

\subsection{Data Preprocessing}

\subsubsection{Data Cleaning and Dataset Splitting}
First, missing values in indoor environmental and meteorological data were filled up using linear interpolation. Then, invalid readings in indoor environmental data were removed, followed by downsampling through averaging to a time granularity of one hour. Subsequently, the dataset was partitioned into three subsets chronologically: a training set for learning the parameters of models, a validation set for fine-tuning hyperparameters and preventing overfitting, and a test set for evaluating the performance of models. The dataset splitting approximately adheres to the empirical ratio of 80:10:10. The three subsets do not overlap in time, avoiding information leakage from the future. We did not identify and address outliers in meteorological data since the provider has validated the data.

\subsubsection{Feature Preparation}
Four temporal features are extracted from timestamps: two binary and two cyclical variables to capture variations in building operations and occupancy. The binary variables include \textit{is holiday} to indicate if a day is a Swedish public holiday and \textit{is weekend} to indicate if a day is a weekend. The cyclical variables are \textit{hour} (integer value from 0 to 23) and \textit{weekday} (integer value from 0 to 6, each value represents a day in a week, starting from Monday).

\subsubsection{Data Transformation}
For observations of target variables indoor temperature and relative humidity over the lookback window, as well as meteorological features such as dry-bulb temperature, relative humidity, dew point temperature, wind speed, and global irradiance, a min-max normalization was performed to scale each of them to a range of $[0, 1]$. All min-max scalers were fitted on the training set and then used for transforming validation and test sets. Cyclical features \textit{hour}, \textit{weekday}, and wind direction were transformed into two dimensions using a sine-cosine transformation.

\subsection{Experimental Setup}

Models were developed to forecast indoor temperature and relative humidity in the next 24 hours. A lookback window size of 168 was used. One baseline model was created based on the seasonal na\"ive (SN) method~\cite{hyndman_forecasting_2018}, namely SN-24. The SN-24 model seeks to use daily seasonality, and each forecast of a target variable is set to the value observed 24 hours ago. 

Five deep learning models (LSTM, TCN, TFT, N-HiTS, and TiDE) were trained to make forecasts. Models were built using the Python programming language (v3.8.16) as well as software packages PyTorch (v1.12.0), darts (v0.27.1), and scikit-learn (v1.2.1). All models were trained on a PC with an NVIDIA GeForce GTX 1080 graphics card. Each model was trained for a maximum of 100 epochs. To prevent overfitting, an early-stopping training strategy was adopted to terminate the training process when the accuracy on the validation set stopped improving after 30 epochs. The models were then evaluated on the edge device equipped in our prototype sensor box, featuring a Quad-core Cortex-A72 64-bit SoC clocked at 1.8GHz, 1GB of RAM, and running on Debian GNU/Linux 12 operating system.

\section{Results and Discussion}
\label{sec:results_and_discussion}

First, results are quantitatively analyzed based on predefined metrics. Then, the results are qualitatively discussed through the exploratory data analysis approach.

\subsection{Quantitative Analysis}
\subsubsection{Predictability of Temperature and Relative Humidity}

As shown in Table~\ref{tab:prediction_accuracy}, both temperature and relative humidity exhibited strong daily seasonality in the test set since the SN-24 model obtained very low CV-RMSE and NMBE. This robust daily seasonality can be attributed to massive walls of Löfstad Castle, acting as a thermal buffer and thereby contributing to the stabilization of indoor climate. More stable temperature and relative humidity make them easier to predict. In all rooms except Room 103, the temperature was more challenging to predict than relative humidity. Most models obtained a significantly larger CV-RMSE when predicting temperature. This phenomenon is due to the fact that the test set data was collected in winter. The change in relative humidity due to the same temperature shift is less pronounced in the lower temperature range due to the exponential relationship between water vapor pressure and temperature.

\begin{table}[!tb]
\centering
\addtolength{\tabcolsep}{-3.5pt}
\caption{The prediction accuracy of different models on the test set. For both metrics of CV-RMSE and NMBE, closer values to zero indicate better performance. Temperature is abbreviated as T, and relative humidity is abbreviated as RH.}
\label{tab:prediction_accuracy}
\begin{tabular}{@{}llrrrrrrrrrrr@{}}
\toprule
\multirow{2}{*}{} & \multirow{2}{*}{Model} & \multicolumn{2}{c}{Room 05} & \multicolumn{1}{c}{} & \multicolumn{2}{c}{Room 3} & \multicolumn{1}{c}{} & \multicolumn{2}{c}{Room 103} & \multicolumn{1}{c}{} & \multicolumn{2}{c}{Room 205} \\ \cmidrule(lr){3-4} \cmidrule(lr){6-7} \cmidrule(lr){9-10} \cmidrule(l){12-13} 
 &  & \multicolumn{1}{c}{T} & \multicolumn{1}{c}{RH} & \multicolumn{1}{c}{} & \multicolumn{1}{c}{T} & \multicolumn{1}{c}{RH} & \multicolumn{1}{c}{} & \multicolumn{1}{c}{T} & \multicolumn{1}{c}{RH} & \multicolumn{1}{c}{} & \multicolumn{1}{c}{T} & \multicolumn{1}{c}{RH} \\ \midrule
\multicolumn{2}{l}{\textit{\textbf{CV-RMSE (\%)}}} & & & & & & & & & & &  \\
 & SN-24 & 2.7 & 2.4 &  & 10.0 & 9.4 &  & 4.7 & 11.5 &  & 20.3 & 5.5 \\
 & LSTM & 3.8 & 1.9 & \textbf{} & \textbf{7.8} & 6.0 &  & 3.7 & 8.2 &  & 22.7 & 7.1 \\
 & TCN & 3.5 & 3.3 &  & 9.8 & 11.7 &  & 4.2 & 14.5 &  & 24.3 & 8.3 \\
 & TFT & 2.3 & 2.3 &  & 13.8 & 8.2 &  & 4.6 & 8.7 &  & 20.7 & 6.5 \\
 & N-HiTS & 6.9 & 2.1 &  & 21.4 & 10.0 &  & 5.9 & 11.7 &  & 27.5 & 6.8 \\
 & TiDE & \textbf{2.1} & \textbf{1.6} & \textbf{} & 9.4 & \textbf{5.5} & \textbf{} & \textbf{3.5} & \textbf{8.1} & \textbf{} & \textbf{11.3} & \textbf{4.8} \\
\multicolumn{2}{l}{\textit{\textbf{NMBE (\%)}}} & \textbf{} & \textbf{} & \textbf{} & \textbf{} & \textbf{} & \textbf{} & \textbf{} & \textbf{} & \textbf{} & \textbf{} & \textbf{} \\
 & SN-24 & 0.8 & $-$0.1 &  & 0.8 & 0.1 &  & 0.0 & $-$0.2 &  & 1.2 & $-$0.4 \\
 & LSTM & $-$1.7 & $-$0.8 &  & $-$2.7 & 1.0 &  & $-$0.9 & $-$2.3 &  & $-$3.3 & 0.7 \\
 & TCN & $-$2.0 & $-$1.8 &  & $-$0.7 & $-$2.1 &  & $-$0.1 & $-$1.5 &  & 1.5 & 0.4 \\
 & TFT & 1.2 & $-$1.3 &  & $-$2.5 & 4.0 &  & $-$1.9 & $-$2.0 &  & 8.6 & 3.9 \\
 & N-HiTS & $-$6.2 & $-$0.9 &  & $-$13.3 & 5.9 &  & $-$3.0 & $-$3.8 &  & $-$3.7 & 1.5 \\
 & TiDE & $-$0.2 & $-$0.3 &  & $-$2.6 & 0.6 &  & $-$0.1 & 0.1 &  & 0.1 & $-$0.7 \\ \bottomrule
\end{tabular}
\end{table}

Among the five deep learning models, only the TiDE model surpassed the performance of the SN-24 model in predicting both temperature and relative humidity of the four rooms. Additionally, the LSTM model demonstrated superior performance with the lowest CV-RMSE when predicting the temperature of Room 3. The high prediction accuracy achieved with the TiDE model offers valuable insights. For scenes like Löfstad Castle, where the indoor environment changes slowly, a simple model might be more effective in extracting patterns.

\subsubsection{The Impact of Heating}

The presence of heating infrastructure on the first floor simplifies the temperature forecasting in Room 103. In contrast to Room 3 and Room 205, all models obtained a lower CV-RMSE when predicting temperature in Room 103. Among unheated rooms, those situated on a higher floor had a lower temperature. As a result, even minor temperature variations lead to a larger CV-RMSE, underscoring the challenge of accurately forecasting temperatures in rooms located on higher floors.

\subsubsection{Computational Cost}

More complicated models generally consume more time to make inferences. As demonstrated in Table~\ref{tab:inference_time}, TCN and TiDE models consumed less inference time due to their ability to process the input data in parallel. LSTM model processes data sequentially, resulting in a slower inference process. The computational cost of the N-HiTS model is slightly higher compared to the LSTM model. The TFT model, which utilizes self-attention mechanisms, requires quadratic time complexity concerning the length of the input sequence, making it less efficient than other models for processing long input sequences.

\begin{table}[!tb]
\centering
\caption{The inference time (mean$\pm$standard deviation) of different models in milliseconds (ms).}
\label{tab:inference_time}
\begin{tabular}{@{}llllll@{}}
\toprule
 & \multirow{2}{*}{Model} & \multicolumn{4}{c}{Inference time (ms)} \\ \cmidrule(l){3-6} 
 &  & Room 05 & Room 3 & Room 103 & Room 205 \\ \midrule
\multicolumn{2}{l}{\textit{\textbf{Temperature}}} &  &  &  &  \\
 & LSTM & 308$\pm$6 & 319$\pm$7 & 326$\pm$6 & 330$\pm$5 \\
 & TCN & \textbf{245$\pm$11} & \textbf{256$\pm$8} & \textbf{262$\pm$9} & \textbf{267$\pm$9} \\
 & TFT & 450$\pm$27 & 483$\pm$28 & 482$\pm$24 & 486$\pm$36 \\
 & N-HiTS & 356$\pm$19 & 395$\pm$18 & 409$\pm$16 & 452$\pm$79 \\
 & TiDE & 274$\pm$9 & 285$\pm$13 & 308$\pm$60 & 297$\pm$6 \\
\multicolumn{2}{l}{\textit{\textbf{Relative humidity}}} & \textbf{} & \textbf{} & \textbf{} & \textbf{} \\
 & LSTM & 312$\pm$5 & 321$\pm$5 & 326$\pm$5 & 335$\pm$15 \\
 & TCN & \textbf{246$\pm$7} & \textbf{261$\pm$13} & \textbf{264$\pm$8} & \textbf{269$\pm$8} \\
 & TFT & 486$\pm$23 & 494$\pm$24 & 489$\pm$25 & 528$\pm$59 \\
 & N-HiTS & 381$\pm$15 & 409$\pm$22 & 426$\pm$24 & 441$\pm$32 \\
 & TiDE & 286$\pm$7 & 298$\pm$15 & 316$\pm$54 & 310$\pm$4 \\ \bottomrule
\end{tabular}
\end{table}

\subsection{Qualitative Analysis}

\begin{figure}[!tb]
\includegraphics[width=\columnwidth]{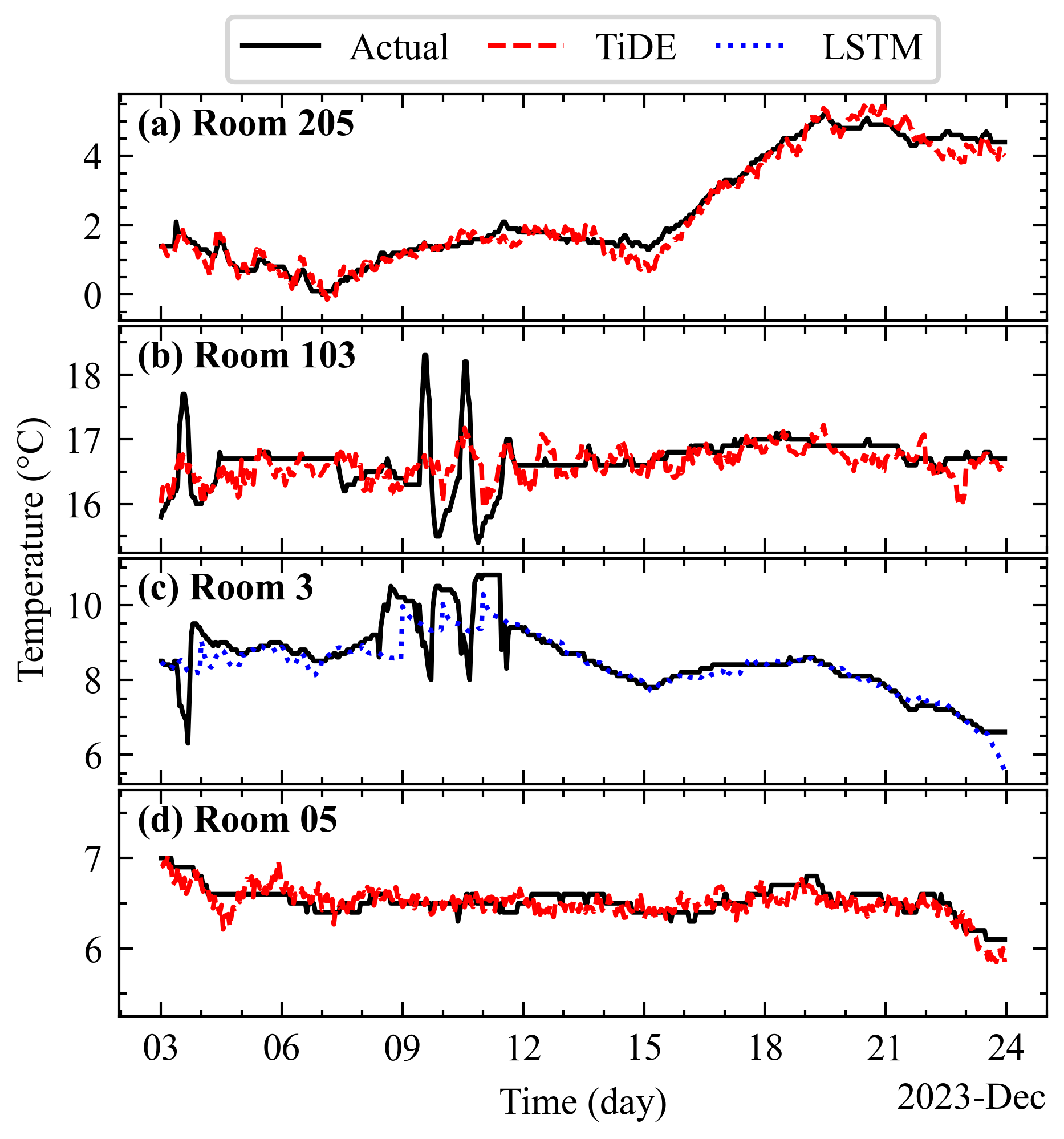}
\centering
\caption{The actual and predicted hourly temperature in (\textbf{a}) Room 205, (\textbf{b}) Room 103, (\textbf{c}) Room 3, and (\textbf{d}) Room 05, of Löfstad Castle from December 3 to December 23, 2023.}\label{fig:fig6}
\end{figure}

\begin{figure}[!tb]
\includegraphics[width=\columnwidth]{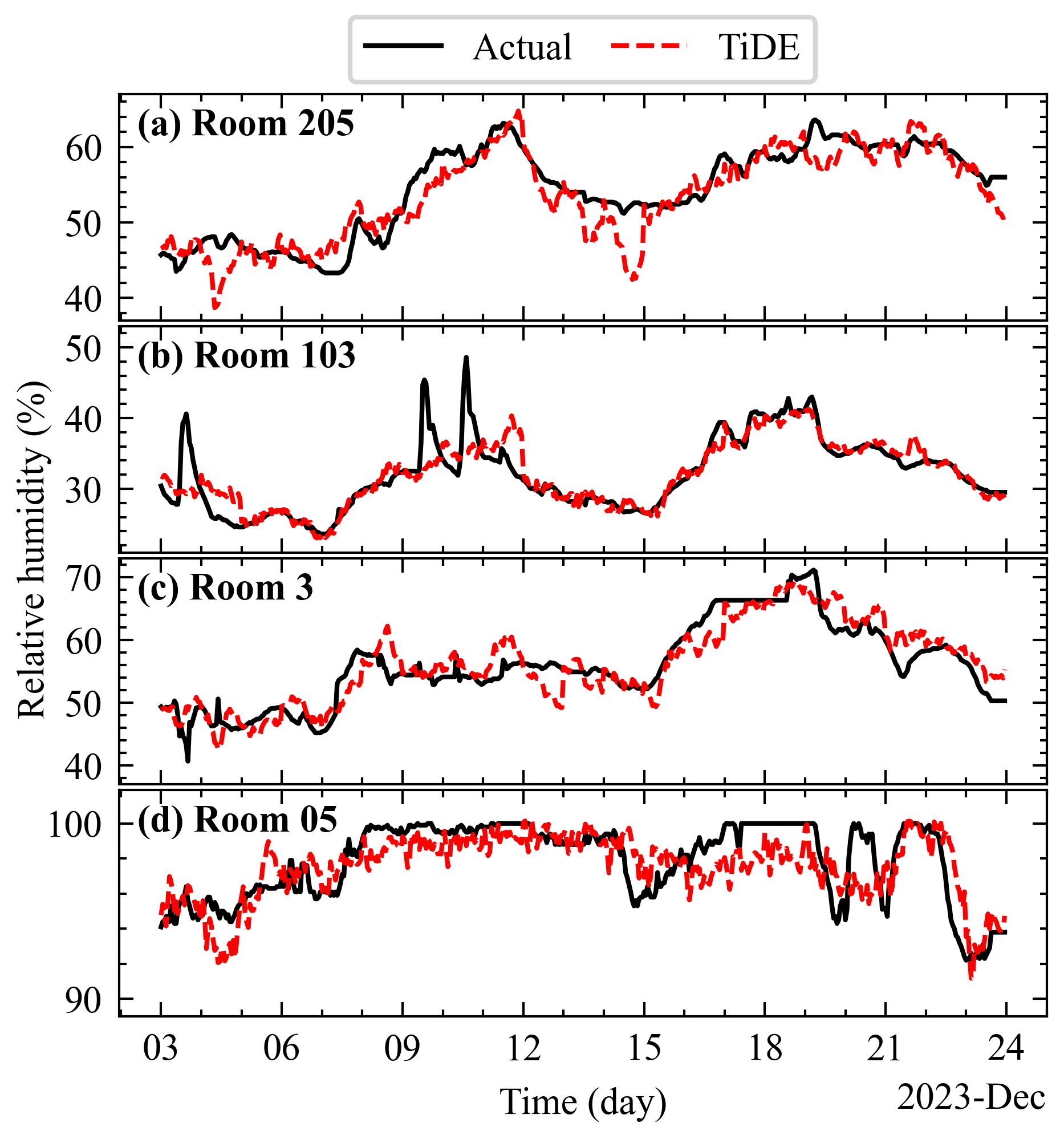}
\centering
\caption{The actual and predicted hourly relative humidity in (\textbf{a}) Room 205, (\textbf{b}) Room 103, (\textbf{c}) Room 3, and (\textbf{d}) Room 05, of Löfstad Castle from December 3 to December 23, 2023.}\label{fig:fig7}
\end{figure}

Previous quantitative analyses have demonstrated that temperature and relative humidity exhibit considerable daily seasonality, resulting in their high predictability. As depicted in Fig.~\ref{fig:fig6} and Fig.~\ref{fig:fig7}, it is evident that both temperature and relative humidity generally exhibit small changes between consecutive days. For instance, in Room 05 (see Fig.~\ref{fig:fig6}{d}), the temperature remained between 6--7~$^{\circ}$C while the relative humidity was mostly higher than 95\% throughout the 21 days. The temperature in Room 103, as shown in Fig.~\ref{fig:fig6}{b}, consistently kept at approximately 16.5~$^{\circ}$C due to the operation of electric radiant heating.

\begin{figure}[!tb]
\includegraphics[width=\columnwidth]{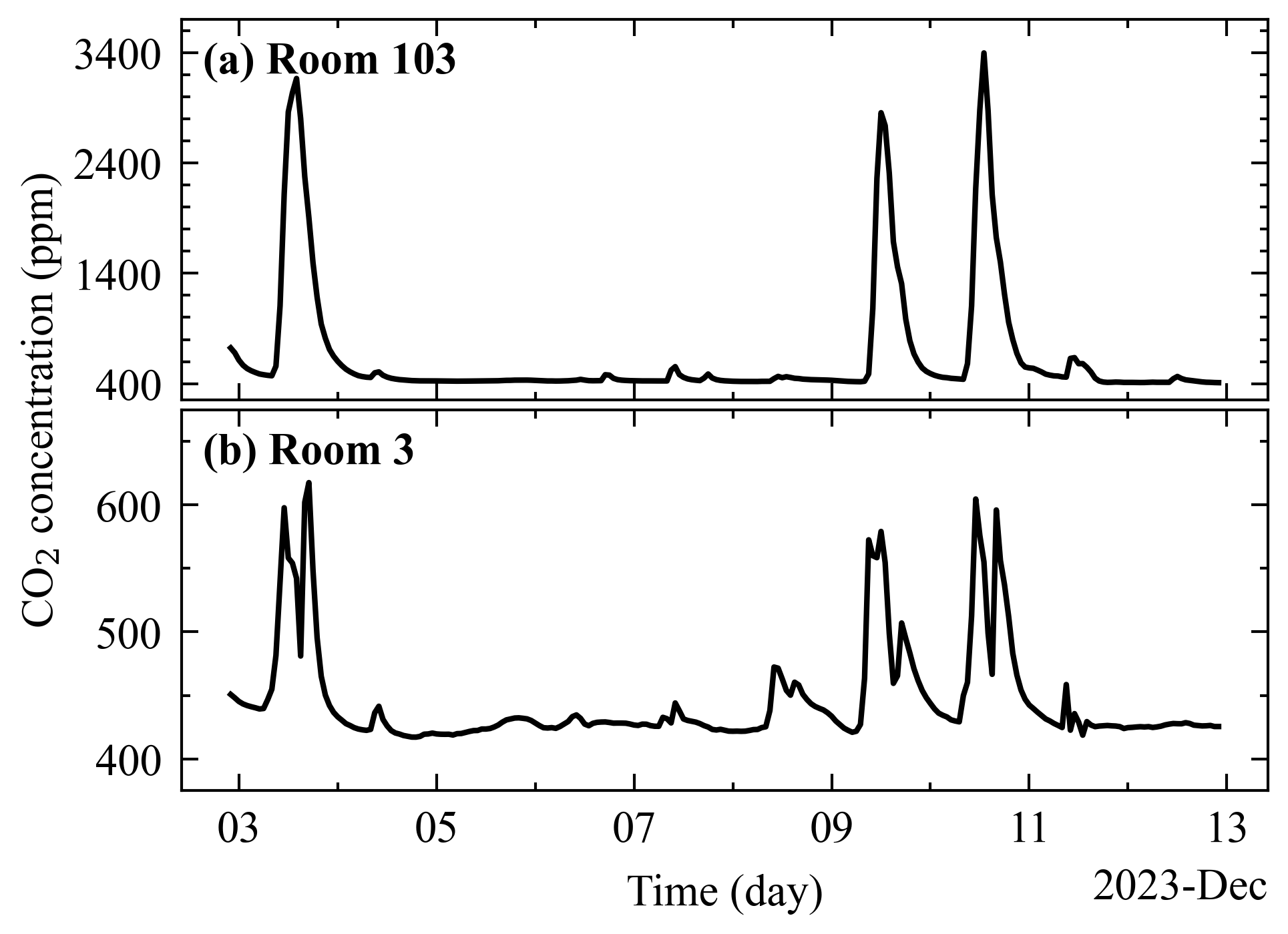}
\centering
\caption{Historical hourly CO\textsubscript{2} concentration in (\textbf{a}) Room 103 and (\textbf{b}) Room 3 of Löfstad Castle from December 3 to December 12, 2023.}\label{fig:fig8}
\end{figure}

Significant variations in the temperature (Fig.~\ref{fig:fig6}{b}) and relative humidity (Fig.~\ref{fig:fig7}{b}) in Room 103 as well as the temperature (Fig.~\ref{fig:fig6}{d}) in Room 3 were observed on December 3rd, 9th, and 10th. Prediction accuracy of models declined during these days. During these occurrences, Room 103 demonstrated a clear pattern of temperature increase succeeded by a decrease, but Room 3 exhibited the contrary trend with a decrease followed by an increase. These notable fluctuations can be attributed to the presence of occupants. The presence of occupants can be confirmed by changes in CO\textsubscript{2} concentration as illustrated in Fig.~\ref{fig:fig8}. On December 3rd, 9th, and 10th, the CO\textsubscript{2} concentration in both rooms experienced a significant rise in comparison to the typical levels. The CO\textsubscript{2} concentration in Room 103 even reached a peak of about 3000 ppm, indicating that there was a gathering of people or other high occupancy in that room during the specified three days. The presence of a large group of people led to gains in both heat and moisture, resulting in the reported rise in temperature and relative humidity in Room 103.

Although the CO\textsubscript{2} concentration in Room 3 increased in these three days, it was not as pronounced as in Room 103. Considering the alternating temperature changes in Room 3, where it first decreased and then increased, a possible explanation is that the door might be left open during the visiting daytime of these days. The open door caused more air to be exchanged between the ground floor and the outside, leading to a temperature decrease in Room 3. After the visitors left and the door closed, temperature in Room 3 increased.

Therefore, to further improve the prediction accuracy of models, it is suggested to include occupancy-related features if they are available. This inclusion enables a more effective consideration of the influence of human behavior on indoor environments, leading to more accurate predictions.

\section{Conclusion}
\label{sec:conclusion}

This paper has presented a solution that combines edge computing, digital twins, and deep learning technology to enhance the understanding of building indoor climate. Adopting ontology in creating parametric digital twins ensures a uniform data representation across different subsystems. The data obtained by digital twins were analyzed using deep learning methods to detect patterns in indoor climate. The findings from a case study in a public historic building in Östergötland, Sweden, have demonstrated that the deep learning architecture time-series dense encoder had robust capabilities in capturing trends with low computational cost. The solution offers facility managers an enhanced understanding of building indoor climate. Therefore, facilities managers can proactively improve maintenance with more precise control of indoor environment. Ultimately, this could lead to cost savings, enhanced human comfort, and a more environmentally friendly built environment. The proposed solution can also be adapted to manufacturing processes to enable proactive maintenance and efficient equipment operations.

\section*{Acknowledgment}

The authors thank Per Stålebro, Emma Vilhelmsson, and Morgan Brålin at the Museum of Östergötland for providing access to Löfstad Castle, offering floorplans, as well as helping install local devices. Gustav Knutsson at Linköping University ordered hardware components, prepared prototype devices, and helped deploy the prototype devices. The Swedish Meteorological and Hydrological Institute is acknowledged for providing meteorological data.

\bibliographystyle{IEEEtran}
\bibliography{references}

\end{document}